\documentclass[aps,prl,reprint,count1]{revtex4-2}

\usepackage{graphicx}% Include figure files
\usepackage{dcolumn} % Align table columns on decimal point
\usepackage{multirow}
\usepackage{colordvi}
\usepackage{color}
\usepackage{epstopdf}
\usepackage{amssymb} 
\usepackage{url}
\graphicspath{{ps}}
\usepackage{hyperref}
\usepackage{tabularx}
\usepackage{amsmath}
\usepackage{cleveref}
\usepackage{graphicx}
\usepackage{makecell}
\usepackage{tikz}
\usepackage{titlesec}
\usepackage{bm}
\usepackage{hyperref}
\usepackage{cleveref}
\usepackage[mathlines]{lineno}
\usepackage{chngcntr}
\usepackage{xspace}
\usepackage{orcidlink}
\input belle2sym.tex
\def\Dbar {\kern 0.2em\bar{\kern -0.2em D}{}\xspace}
\def\Dzb {\ensuremath{\Dbar^0}\xspace}

\crefname{figure}{Fig.}{Figs.} 

\Crefname{figure}{Figure}{Figures} 
\begin{document}

\preprint{APS/123-QED}

\begin{minipage}[h]{0.45\textwidth}
\begin{flushright}
% Internal paper draft details
% BELLE2-PUB-DRAFT-2024-007\_V22 \\
% Note: BELLE2-NOTE-PH-2023-048 \\
% Authors: Daniel Marcantonio, Phillip Urquijo
% Intended journal: PRL \\
% Preprint info
Belle II Preprint 2025-029 \\
KEK Preprint 2025-36 \\     
\end{flushright}
\end{minipage}
\vspace{20pt}
\texorpdfstring{}{}
\title{\texorpdfstring{A search for feebly-interacting particles in $B$ decays with missing energy at Belle}{A search for feebly interacting particles in missing energy B decays at Belle}}

%%% Paper:    Feebly interacting particles
%%% Journal:  PhysRev
%%% Contacts: D. Marcantonio, P. Urquijo
%%% ====================================================================
%%% Use \input{pub093-orcid} to insert this material into your latex file.
  \author{M.~Abumusabh\,\orcidlink{0009-0004-1031-5425}} % 26883
  \author{I.~Adachi\,\orcidlink{0000-0003-2287-0173}} % 2590
% \author{K.~Adamczyk\,\orcidlink{0000-0001-6208-0876}} % 2239
  \author{L.~Aggarwal\,\orcidlink{0000-0002-0909-7537}} % 10083
% \author{P.~Ahlburg\,\orcidlink{0000-0002-9832-7604}} % 2367
  \author{H.~Ahmed\,\orcidlink{0000-0003-3976-7498}} % 11323
% \author{J.~K.~Ahn\,\orcidlink{0000-0002-5795-2243}} % 7423
  \author{Y.~Ahn\,\orcidlink{0000-0001-6820-0576}} % 14363
  \author{H.~Aihara\,\orcidlink{0000-0002-1907-5964}} % 2223
  \author{N.~Akopov\,\orcidlink{0000-0002-4425-2096}} % 9443
  \author{S.~Alghamdi\,\orcidlink{0000-0001-7609-112X}} % 27804
  \author{M.~Alhakami\,\orcidlink{0000-0002-2234-8628}} % 28103
  \author{A.~Aloisio\,\orcidlink{0000-0002-3883-6693}} % 2194
% \author{S.~Al~Said\,\orcidlink{0000-0002-4895-3869}} % 6823
% \author{A.~Alsharari\,\orcidlink{0000-0002-6993-1597}} % 27803
% ...
% 13403
% \author{S.~Zou\,\orcidlink{0000-0003-3377-7222}} % 19363
\collaboration{The Belle and Belle II Collaborations}

\collaboration{Belle and Belle II Collaborations}

\begin{abstract}
We present a search for an invisible hidden-sector particle $\Xinv$, produced in $\Bz \to \Dzb\Xinv$ and $\Bpm\to h\Xinv$ decays, where $h = \pipm$, \Kpm, $D_{s}^\pm$, $p^\pm$. The search is performed using $\ep\en$ collision data recorded with the Belle detector, corresponding to 711 $\invfb$. No significant signal is observed. We set 90\% confidence level upper limits ranging between $10^{-4}$ and $10^{-6}$ on the branching fraction $\mathcal{B}(B \to h\Xinv)$ as a function of $m_{\Xinv}$. Corresponding limits are set on $\mathcal{B}(B \to h\X)\times\mathcal{B}(\X\to\gamma\gamma)$ for lifetimes $c\tau_\X$ between 10~$\mu$m and 10~m. Many of these limits are the first direct constraints on their respective decays. Our results provide the most stringent exclusion limits to date on the branching fractions for all search channels, and exclude previously unexplored regions of parameter space relevant to several new physics models. 
\end{abstract}

\maketitle
% Introduction
In recent years, the absence of evidence for physics beyond the standard model (SM) at the energy frontier has shifted attention toward searches for new particles with masses comparable to those of known SM particles. If such particles exist and have so far avoided detection, this would likely be due to their small couplings to SM particles; such hypothetical particles are termed feebly-interacting particles (FIPs)~\cite{Agrawal_2021}. The interactions of FIPs with SM particles can be governed by low-dimensional operators, or \textit{portals}, classified according to mediator spin \cite{Batell:2022hko}. This phenomenological framework enables straightforward comparison across experimental searches.

% FIPs
Many particles proposed in extensions to the SM are FIP candidates. Pseudoscalar axion-like particles (ALPs) arise in many extensions to the SM \cite{PhysRevD.16.1791, PhysRevLett.40.223, PhysRevLett.40.279} and offer potential solutions to longstanding problems in physics \cite{Graham_2015}, as potential dark matter candidates \cite{ABBOTT1983133,DINE1983137,PRESKILL1983127} or dark sector mediators \cite{PhysRevD.79.075008}. Constraints on ALP couplings to photons and electrons are already stringent across the MeV-GeV range \cite{Graham_2015, Essig_2013, Marsh_2016,BaBar:2021ich,Sungjin25}. ALPs produced in flavor-changing neutral current (FCNC) processes, such as those in $B$ decays, are especially promising to detect, as these transitions are suppressed in the SM, enhancing sensitivity to physics beyond the SM~\cite{Bauer:2021mvw, zhang2023belle, NA62_2021,NA62:2020xlg, NA62:2023olg, NA62:2020pwi, BNL-E949:2009dza}. These decays depend on the coupling of ALPs to the $\Wpm$ bosons, which, compared to the couplings to photons and electrons is less constrained~\cite{Izaguirre_2017, PhysRevLett.115.221801, Dolan_2015, Dolan_2015Erratum, PhysRevD.83.115009, PhysRevD.79.075008, PhysRevD.71.014015}.

Scalars can also be dark matter candidates \cite{Silveira:1985rk} or dark sector mediators~\cite{Pospelov:2007mp}, and can play roles in solving a range of problems in physics \cite{Patt:2006fw, OConnell:2006rsp, Piazza:2010ye,PhysRevLett.115.221801,Espinosa:1993bs, Profumo:2007wc, Croon:2019ugf, Bezrukov:2009yw}. Dedicated studies in $B$ decays further probe their parameter space \cite{PhysRevD.101.095006, PhysRevD.93.025026}. While less commonly studied in $B$ decays, light vector particles are also in principle allowed \cite{Datta_2023, Fuyuto_2016}, with potential relevance to the excess rate measured for $\Bp\to\Kp\nu\nub$~\cite{gabrielli2024}.

% Invisible particles
Although portals provide a general classification of models, the observability of a new particle depends on its decay properties. A FIP may appear invisible because it is a dark matter candidate, predominantly decays into dark matter candidates, or because it has a long lifetime and decays outside the detector volume. In any case, the result is a missing energy signature.

% In this Letter...
In this Letter, we report on a search for an invisible FIP, $\Xinv$, produced in $\Bz \to \Dzb\Xinv$ and $\Bp\to h\Xinv$ decays, where $h = \pip, \Kp, \Ds, \pp$. Charged-conjugate decays are implied throughout. The data were collected with the Belle detector \cite{Belle:2000cnh} at the energy-asymmetric $\ep\en$ KEKB collider \cite{Kurokawa:2001nw} at a center-of-mass (c.m.)\ energy $\sqrt{s} = 10.58 \gev$. The search is performed on a dataset containing $(770~\pm~11)\times~10^6$\ $\ep\en\to\Y4S\to B\Bb$ events, corresponding to an integrated luminosity of 711 fb$^{-1}$. 

% Experimental part starts here
The search for $\Xinv$ is carried out by scanning for narrow peaks in the momentum of the recoiling hadron in the frame of its parent (signal-side) $B$ meson, denoted $\ph$. The other (tag-side) $B$ meson in the event is also reconstructed to kinematically constrain the event. A scan is performed across the \ph distribution looking for excesses above the expected background. The current best limits on the branching fractions of $\Bp\to\Kp\Xinv$ and $\Bp\to\pip\Xinv$ come from searches for the SM channels $\Bp \to \Kp\nu\nub$ and $\Bp \to \pip\nu\nub$ at $B$-factories, which have excluded branching fractions down to approximately $10^{-5}$ for both channels \cite{Lees_2013, PhysRevD.87.111103, Chen_2007, PhysRevD.96.091101}. Recently, there has been a surge of interest in $b\to s\nu\nub$ transitions, as Belle II reported evidence of $\Bp\to \Kp\nu\nub$ \cite{belleiiBtoKnunu}. The best existing limit on $\mathcal{B}(\Bp \to \pp\Xinv)$ is about $10^{-6}$, set by a recent BABAR search \cite{BABAR_proton}. To date, there have been no dedicated searches for the two decays involving charm mesons discussed in this Letter, making this the first search for $\Bp \to \Ds \Xinv$ and $\Bz \to \Dzb \Xinv$.

% Introduction to X --> γγ 
While our primary focus is on invisible final states, we also assess the sensitivity of our selection to scenarios where a particle $\X$ decays via $\X\to\gamma\gamma$ with decay length comparable to the detector size. Although our analysis is optimized for the invisible case, such visible decays can mimic the $\Xinv$ signature when $\X$ decays outside the sensitive detector volume or when the decay products escape detection.

% Detector description
The Belle detector has a cylindrical symmetry around the beamline, with the $z$-axis defined as the direction of the electron beam. The Belle detector comprises six sub-detectors, listed in the following from innermost to outermost: the silicon vertex detector, the central drift chamber, the aerogel Cherenkov counter, the time-of-flight scintillation counter, the electromagnetic calorimeter (ECL), and the \KL and muon detector. The ECL consists of a cylindrical barrel region surrounding the beamline and forward and backward endcap regions at either end, with the forward endcap located in the positive $z$ direction. A solenoid produces a 1.5~T magnetic field through the five innermost detectors. Further details of the detector are in Ref. \cite{Belle:2000cnh}.

% MC generation
Signal events are simulated using the Monte Carlo (MC) event generator \verb|EvtGen| \cite{Lange:2001uf} for a range of \Xinv masses from 1 MeV/$c^2$ to just below the kinematically allowed upper limit of $m_B - m_h$, for both invisible \Xinv and $\X\to\gamma\gamma$. For the latter, the lifetime $c\tau_X$ is generated in a range from $10\ \mu\rm{m}$ to $10$~m. The background processes $\ep\en\to\Upsilon(4S) \to B\Bb$ (with SM $B$ decays) and $\ep\en \to q\qbar(\gamma)$ (where $q = {u,d,s,c}$) are simulated using \verb|EvtGen| and \verb|PYTHIA| \cite{Sj_strand_2006}, with final-state radiation simulated using \verb|PHOTOS| \cite{Barberio:1993qi}. The detector response is modeled with \verb|GEANT3| \cite{Brun:1987ma}. Both experimental and simulated events are converted to the Belle II format \cite{Gelb_2018}, and then reconstructed and analyzed using the Belle II analysis software framework \cite{Kuhr_2018, the_belle_ii_collaboration_2022_6949513}. To avoid experimenter’s bias, the selection criteria and analysis procedure are finalized using simulated events before examining the data.

% Btag reconstruction
The tag-side $B$ meson is reconstructed from hadronic decay modes using the full event interpretation (FEI) algorithm \cite{Keck_2019}. It must satisfy $\Mbc \equiv \sqrt{s/4 - |p^{*}_{B}|^2} > 5.27$ \gevcc and $|\Delta E| \equiv |E^*_B - \sqrt{s}/2| < 0.1$ \gev, where $p^*_B$ and $E^*_B$ are the momentum and energy of the $B$ meson. Here and throughout the paper, quantities in the $\ep\en$ c.m. frame are indicated by an asterisk. The signal quality returned by the FEI is required to exceed 0.005. 

% Reconstruction and basic event selection
All signal channels contain at least one charged particle in the final state. We select reconstructed trajectories of charged particles (tracks) with a point of closest approach to the interaction point (IP) in the radial direction satisfying $dr <$~2~cm, and in the $z$-direction satisfying $|dz| <$~ 3~cm. All final-state particles are required to have a momentum $p^* > 0.1$ \gevc. 

Charged particles are identified using a binary ratio $P$ of likelihoods $\mathcal{L}$, which compares two particle hypotheses $i$ and $j$: $P_{ij} \equiv \mathcal{L}_i / (\mathcal{L}_i + \mathcal{L}_j)$ \cite{Nakano:2002jw}. In the search for $\ckaon$ ($\cpion$), we select charged kaons (pions) with 88\% (92\%) efficiency with a 4\% (7\%) pion (kaon) mis-identification rate. In the search for $\cproton$, protons are identified with 97\% efficiency with 18\% and $<1\%$ kaon and electron mis-identification rates, respectively. 

In the searches for $\cDs$ and $\cDz$, the $\Ds$ and $\Dz$ mesons are reconstructed from specific final states: $\Ds\to \Kp\Km\pip$ and $\Dz \to \Km\pip$, $\Km\pip\piz$, or $\Km\pip\pim\pip$. The $\pip$ and $\Kp$ candidates in these decays must meet the same IP requirements as above, with looser particle identification selections, corresponding to 97\% (98\%) charged kaon (pion) efficiency, with a pion (kaon) mis-identification rate of 18\% (23\%). A vertex fit is performed as detailed in Ref. \cite{Krohn_2020}. The $\Ds$ and $\Dz$ candidates are required to lie within 3 times the mass resolution of their nominal masses \cite{ParticleDataGroup:2024cfk}. 

Neutral pion candidates are reconstructed using two ECL energy deposits (clusters) not matched with tracks, which we identify as photons, with an invariant mass between 118~\mevcc and 150~\mevcc (corresponding to an invariant mass window of approximately $\pm2\sigma$). This retains 97\% of $\piz$ produced in the decay $\Dzb\to\Kp\pim\piz$. To suppress background in the ECL, we apply minimum photon energy thresholds of $50~\mev$ in the barrel and $100~\mev$ in the endcaps.

% ROE
The rest of event (ROE) refers to all reconstructed charged particles and photons in the event that are not associated with the signal- or tag-side decays. Charged particles are only considered in the ROE if $dr <$~2~cm and $|dz| <$~5~cm, and photons are only considered if their energies are greater than 50~\mev, 100~\mev, and 150~\mev in the barrel, forward endcap, and backward endcap, respectively. Events are retained only if there are no charged particles in the ROE and the total energy from photons in the ROE ($E_{\rm{ECL}}$) does not exceed 1.5 \gev.

% Further background suppression
To further suppress background processes, we train boosted decision trees (BDTs) \cite{Chen:2016:XST:2939672.2939785} on simulated events. These classifiers are trained using a set of variables describing the event topology and kinematics: the ratio of the second to zeroth Fox-Wolfram moment ($R_2$) \cite{PhysRevLett.41.1581}, the modified moments described in \cite{ABE2001151,PhysRevLett.91.261801}, thrust-related quantities ($\cos\theta_T$, $\cos\theta_{T,z}$, $\hat{T}(O)$, and $\hat{T}(B)$), the nine CLEO cones \cite{PhysRevD.53.1039}, the \Btag energy difference and flight distance, the missing mass squared, and \eecl \cite{Bevan_2014,kouBelleIIPhysics2019}. 

The first classifier, BDT1, is trained using a simulated dataset equivalent to 400 \invfb to distinguish events with $B$ decays from $\ep\en\to q\qbar$ events. Using the same variables, a second classifier, BDT2, is then trained on signal MC events and background events that pass a 90\% signal efficiency cut on the output of BDT1. This training uses a dataset equivalent to about 5~\invab.

% Selection criteria optimisation
The selection criteria for the two classifier outputs are optimized using a two-dimensional grid search, with a Punzi figure of merit \cite{Punzi:2003bu} with a target statistical significance of $3\sigma$. The selection criteria for the classifier outputs are optimized for each channel individually. Averaged across all channels, a signal efficiency of 84\% and background rejection of 86\% are achieved. In simulated signal events, the average number of candidates per event is 1.5 for $\Bp\to h\Xinv$ and 1.9 for $\Bz\to\Dzb\Xinv$. 
In cases where multiple candidates satisfy the selection criteria, the candidate with the highest FEI signal probability is retained. 
The correctly reconstructed candidate is selected more than $97\%$ of the time for $\Bp\to h\Xinv$, and $80\%$ of the time for $\Bz\to\Dzb\Xinv$.

% Validation
We verify that the BDT classifiers perform similarly in data and simulation by studying two-body SM $B$ decays that are kinematically similar to the signal processes. For $h = \Kp$, we examine a sample with two ROE tracks near the $\jpsi\to\ell^+\ell^-$ (where $\ell = e, \mu$) peak arising from $\Bp \to \jpsi \Kp$ decays. For $h = \Dz$, we examine an \eecl sideband with $2.25 < \eecl < 3.5$ \gev near the $\piz\to\gamma\gamma$ peak arising from $\Bz\to\Dzb\piz$ decays. Classifiers' responses for these channels are validated in these respective sidebands. The BDT output distributions show good agreement between data and simulation. However, the FEI efficiency exhibits significant differences arising from imperfect modeling of detector response, particle identification, and the branching fractions of the numerous decay channels used in the tag reconstruction. 

\begin{figure}
 \centering
 \includegraphics[width=\linewidth]{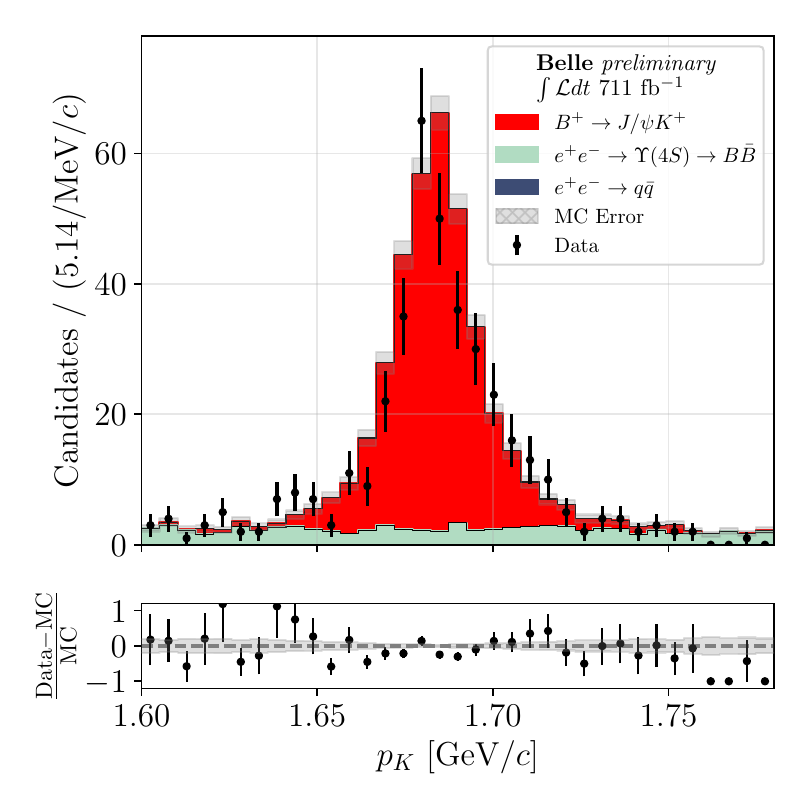}
 \includegraphics[width=\linewidth]{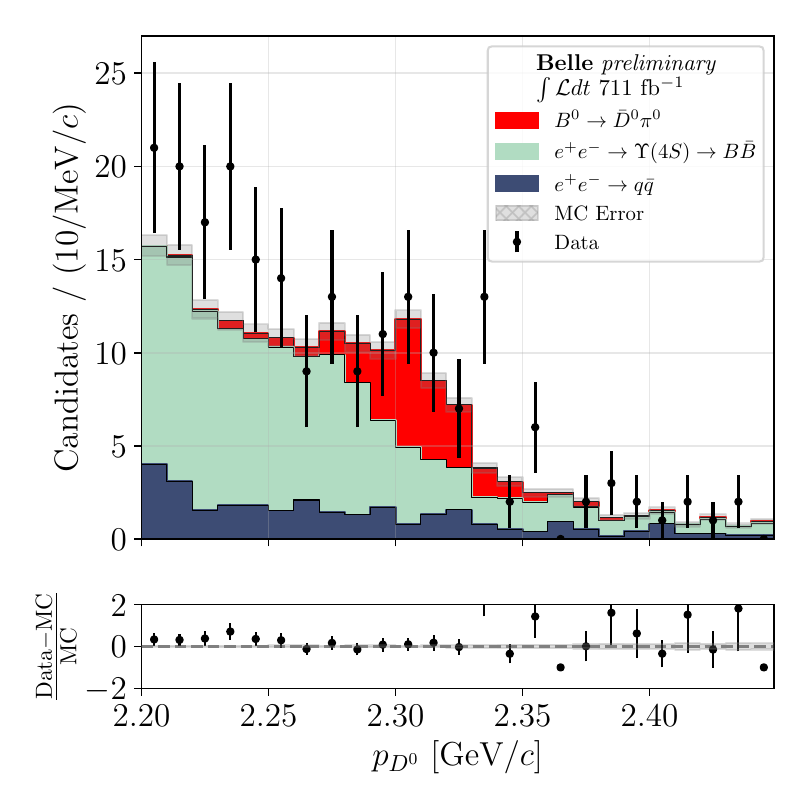}
 \caption{\label{fig:control_modes} Momentum distributions of the $\Kp$ and $\Dz$ in the $\Bsig$-frame for the two control channels, $\Bp\to\jpsi\Kp$ (top) and $\Bz\to\Dzb\piz$ (bottom), respectively, overlaid with the expected distributions from simulation.}
\end{figure}

% FEI calibration
The FEI efficiency is calibrated using $B\to D^{(*)}\ell\nu$ decays \cite{Belle:2022yzd}. The correction is taken as the ratio of signal yields between real and simulated data; the values are $0.79 \pm 0.02$ and $0.78 \pm 0.02$ for charged and neutral $B$ mesons, respectively. These values are cross-checked using sidebands containing $\Bp\to\jpsi\Kp$ and $\Bz\to\Dzb\piz$ events for charged and neutral $B$ decays, respectively. In the two-track sample, we fully reconstruct the $\Bsig$ decay $\Bp\to\jpsi\Kp$, requiring $|\Delta E|<0.1$~GeV, and only keep candidates with $1.6 < p_K < 1.78~\gevc$. No additional selection is applied to the $\Bz\to\Dzb\piz$ sideband, in which candidates are kept with $2.20 < p_{\Dz} < 2.45 \gevc$. The experimental and simulated momentum distributions for both control modes are shown in \cref{fig:control_modes}. We perform a fit to the \Kp and \Dzb momentum distributions. The peaks in these distributions, corresponding to the recoiling $\jpsi$ and $\piz$, are modeled using a sum of two Gaussian functions. The background is modeled with a third-order polynomial. The signal peak width is allowed to float in the fit to data. The width scale factor is used to correct for resolution differences between simulation and experimental data, and is found to be $1.21\pm0.05$ and $1.08\pm0.23$ for $\Bp\to\jpsi\Kp$ and $\Bz\to\Dzb\piz$, respectively. The ratios of signal yields between experimental and simulated data are $0.75\pm0.09$ and $0.84\pm0.34$ for charged and neutral $B$ mesons, respectively. These values agree with the ratios derived from $B\to D^{(*)}\ell\nu$ decays.

% Selection efficiency
The signal selection efficiency depends on the decay mode: for $h = \pip,\Kp,\pp$, it is around 0.5\%, while for $h=\Ds,\Dzb$, it is around 0.01\%, as only a fraction of the $\Ds$ and $\Dz$ decays are reconstructed. The efficiency is mostly flat as a function of $m_{\Xinv}$, but reduces as $m_{\Xinv}$ approaches the kinematic limit of $m_B - m_h$.

% Fit approach
The signal yield is extracted by performing extended, unbinned maximum likelihood fits to $\ph$. The fit model consists of signal and background components. The signal component is modeled as a sum of two Gaussian functions with a shared mean $\mu$ and different widths $\sigma_1$ and $\sigma_2$. The fit is first performed on simulated signal events, allowing the widths and relative contributions of the Gaussian components to float. Subsequently, these fit results are parameterized as a function of the $\ph$ value corresponding to the generated $m_{\Xinv}$ of the sample. The resolution $\sigma$ of the peak is the quadrature sum of the two width parameters, taking into account their relative weights, and the width scaling factor derived from the control mode fits. This approach provides a well-defined signal probability density function (PDF) at any scan point. 

To parameterize the background, we use a kernel density estimator (KDE) \cite{Cranmer_2001}, with the kernel width set by the local event density. The KDE is built from simulated background events and has a fixed shape during the signal extraction fit. Differences in the $\ph$ distribution between off-resonance data and simulated $\ep\en\to q\qbar$ events are accounted for by applying a linear correction to the $\ep\en\to q\qbar$ background PDF. Independent fits are conducted for several hundred mass hypotheses for each channel, with scan steps equal to half the signal resolution. The resolution ranges from about 5 MeV/$c$ to 40 MeV/$c$, generally increasing as $m_{\Xinv}$ decreases. The KDE for a given fit window is defined over a $\pm15\sigma$ window. A combined signal and background fit is performed in a $\pm10\sigma$ window to mitigate the impact of the KDE boundary problem \cite{Cranmer_2001}.

For all channels, a toy MC study is performed in which simulated events are sampled with varying yields of signal, $q\qbar$, and $B\Bb$ background events. The mean signal yield obtained from the combined fitting method closely matches the number of signal events in the sampled datasets, indicating the signal extraction method responds linearly to the presence of signal events.

% Peaking backgrounds
The only background processes that can mimic the peaking structure of the signal are two-body $B$ decays. For common processes with narrow peaks, such as $\Bp \to \Kp \Dzb$, events within three times the resolution of the peak are vetoed. For some of these processes, such as $\Bp \to \Kp f_0(1370)$, the recoil peak is so broad that it can be safely treated as a non-peaking background. For very rare processes where the expected number of events is below the expected sensitivity of the search, such as $\Bp\to\Kp K^\ast(892)^0$ (which has a branching fraction $\mathcal{O}(10^{-7})$) \cite{LHCb:2014lcy}, a signal-like PDF component corresponding to the SM particle mass is added to the fit, and its yield is allowed to float within $\pm100\%$ of the predicted branching fraction or two events, whichever is greater.

% Systematic uncertainties
The dominant systematic uncertainty is the 2.5\% uncertainty on the FEI correction factor for both charged and neutral $B$ mesons. Other sources include uncertainties in particle identification, tracking efficiency, corrections made to the $\ph$ distributions from $\ep\en\to q\qbar$ events, and limited signal MC statistics. For channels involving $\Ds$ and $\Dzb$ mesons, uncertainties in the branching fractions of the specific decay modes used for their reconstruction are included, with an additional uncertainty from $\piz$ reconstruction in the $\Dzb$ case. Peaking background yields also contribute to the total uncertainty. Statistical uncertainties dominate across all five channels, while total systematic uncertainties remain below a few percent.

The branching fraction $\mathcal{B}(B \to h\Xinv)$ is extracted directly from an extended maximum likelihood fit to the $\ph$ spectrum. The selection efficiency and the number of $\BB$ pairs are included as nuisance parameters, allowing their associated uncertainties to be propagated through the fit.

\begin{figure}[htb]
 \centering
 \includegraphics[width=0.49\textwidth]{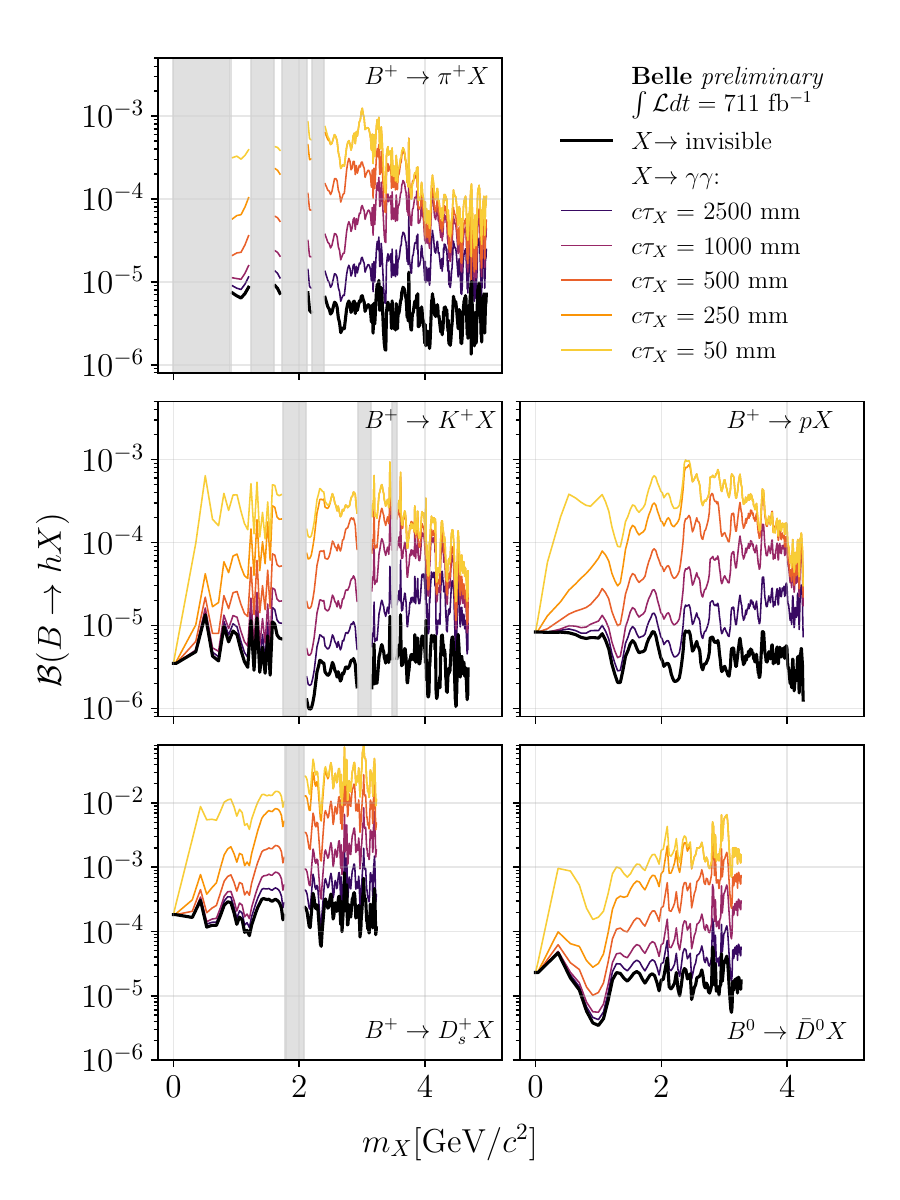}
 \caption{\label{fig:BF_SensAll}Upper limits on the branching fraction $\mathcal{B}(B\to h\Xinv)$ at 90\% CL as a function of $m_{\Xinv}$ are shown in black. Replacing the invisible decay assumption with $\X\to\gamma\gamma$ yields the colored limits, shown for several orders of magnitude of $c\tau_\X$. Mass regions vetoed for known SM particles $\piz$, $K^{(*)0}$, $D^{(*)0}$, $\eta_c$, $\chi_{c1}$, and $\psi(2S)$ are shaded gray. Not all veto regions apply to every channel.}
\end{figure}

% Significance and result extraction
The local significance for each mass hypothesis is defined as $\mathcal{S}_{\rm{local}} = \sqrt{2 \cdot (\log\mathcal{L}_{s+b} - \log\mathcal{L}_{b})}$, where $\mathcal{L}$ denotes the maximum likelihood under the signal-plus-background and background-only hypotheses. Given the scan over several hundred mass points per channel, the look-elsewhere effect \cite{Lyons_2008} is addressed by converting $\mathcal{S}_{\rm{local}}$ to a global significance $\mathcal{S}_{\rm{global}}$ using the trial factor method of Ref.~\cite{Gross_2010}.

No significant excess is observed. The most significant local excess, $\mathcal{S}_{\rm{local}} = 2.95\sigma$, is observed at $m_{\Xinv} = 3.28\ \gevcc$ in the $\Bp \to \pip\Xinv$ channel, corresponding to $\mathcal{S}_{\rm{global}} = 0.65\sigma$. Upper limits at the 90\% confidence level (CL) are set as a function of $m_{\Xinv}$ using the modified frequentist $\mathrm{CL_s}$ method~\cite{CLs}. Limits on the branching fraction are computed at each mass point, as shown in \cref{fig:BF_SensAll}. Notably, the limit on $\mathcal{B}(\Bp\to\Kp\Xinv)$ excludes that decay as an explanation for the observed excess in $\Bp\to\Kp\nu\nub$ \cite{Belle-II:2025lfq} for the values of $m_{\Xinv}$ considered. However, the $\Bp\to\Kp\nu\nub$ excess could arise in mass regions vetoed by this work \cite{PhysRevD.109.115006}.

% Long-lived particle results 
Searches for long-lived particles that decay via $\X\to\gamma\gamma$ are motivated by many ALP models \cite{Agrawal_2021,Izaguirre_2017}. Here, $\X$ refers to a FIP that may not be completely invisible due to a lab-frame decay length on a similar scale as the detector. The sensitivity to such decays is evaluated by applying the nominal $B\to h\Xinv$ selection criteria to the aforementioned simulated signal sample containing $B\to h\X(\to\gamma\gamma)$ decays. We stress that this is not a dedicated search for $\X\to\gamma\gamma$, nor is any alternative reconstruction or statistical treatment performed. Instead, the $\X\to\gamma\gamma$ study serves only to quantify at what lifetimes such decays would mimic the invisible scenario under the nominal selection. These results are shown as the colored lines in \cref{fig:BF_SensAll}. Since the selection criteria are designed to reject events with substantial \eecl, and the decay products of $\X\to\gamma\gamma$ are more likely to be detected in the ECL for smaller values of $c\tau_\X$, the sensitivity is worse at shorter lifetimes. 

For some benchmark masses and lifetimes, the $X$ selection efficiency was recalculated for several other final states: $\ep\en$, $\mup\mun$, $\pip\pim$, $\Kp\Km$, $\pi\pi\eta$, and $\pi\pi\gamma$. Photon pairs, for a given $m_\X$ and $\tau_\X$, are the decay products that deposit the most energy in the ECL and therefore cause the event to fail the \eecl $<1.5~\gev$ requirement. This makes the $\X\to\gamma\gamma$ limits from this work a conservative estimate for any of the other final states listed above, several of which are interesting in the context of hadronically-decaying ALP scenarios \cite{PhysRevLett.123.031803,balkin2025covariantdescriptioninteractionsaxionlike}. 

% Summary
In summary, we present a search for invisible particles $\Xinv$ produced in $\Bp\to h\Xinv$, $h = \pip,\Kp,\pp,\Ds$, and $\Bz\to\Dzb\Xinv$ decays, using the 711~\invfb Belle dataset. No significant excess is observed, and 90\% upper limits are set on the branching fraction $\mathcal{B}(B \to h\Xinv)$ for all channels. We provide another set of limits under the alternative assumption that the new particle decays via $\X\to\gamma\gamma$ and has a lifetime $c\tau_\X$ between 10~$\mu$m and 10~m. These results provide the most stringent branching fraction exclusion limits in these channels to date and are the first search results for the channels $\Bp\to\pip\Xinv$, $\Bp\to \Ds\Xinv$, and $\Bz\to\Dzb\Xinv$. 

The limits in \cref{fig:BF_SensAll} place the strongest constraints on the ALP-$W$ coupling, several ALP-quark couplings, and the dark scalar-Higgs mixing parameter (outlined in the End Matter). Furthermore, the limits on $\Bp\to\pp\Xinv$ provide the best constraints on the $B$-mesogenesis \cite{Elor_2019, Alonso_lvarez_2021} parameter space, and on the $b$ quark coupling to the lightest neutralino in R-parity-violating supersymmetry models \cite{Dib_2023}.

This work, based on data collected using the Belle detector, which was
operated until June 2010, was supported by 
the Ministry of Education, Culture, Sports, Science, and
Technology (MEXT) of Japan, the Japan Society for the 
Promotion of Science (JSPS), and the Tau-Lepton Physics 
Research Center of Nagoya University; 
the Australian Research Council including grants
DP210101900, % Urquijo
DP210102831, % Sevior
DE220100462, % Hsu
LE210100098, % Infrastructure
LE230100085; % Infrastructure
Austrian Federal Ministry of Education, Science and Research (FWF) and
FWF Austrian Science Fund No.~P~31361-N36;
National Key R\&D Program of China under Contract No.~2022YFA1601903,
National Natural Science Foundation of China and research grants
No.~11575017,
No.~11761141009, 
No.~11705209, 
No.~11975076, 
No.~12135005, 
No.~12150004, 
No.~12161141008, 
and
No.~12175041, 
and Shandong Provincial Natural Science Foundation Project ZR2022JQ02;
the Czech Science Foundation Grant No. 22-18469S;
Horizon 2020 ERC Advanced Grant No.~884719 and ERC Starting Grant No.~947006 ``InterLeptons'' (European Union);
the Carl Zeiss Foundation, the Deutsche Forschungsgemeinschaft, the
Excellence Cluster Universe, and the VolkswagenStiftung;
the Department of Atomic Energy (Project Identification No. RTI 4002), the Department of Science and Technology of India,
and the UPES (India) SEED finding programs Nos. UPES/R\&D-SEED-INFRA/17052023/01 and UPES/R\&D-SOE/20062022/06; 
the Istituto Nazionale di Fisica Nucleare of Italy; 
National Research Foundation (NRF) of Korea Grants
No.~2021R1-A6A1A-03043957,
No.~2021R1-F1A-1064008,
No.~2022R1-A2C-1003993,
No.~2022R1-A2C-1092335,
No.~RS-2016-NR017151,
No.~RS-2018-NR031074,
No.~RS-2021-NR060129,
No.~RS-2023-00208693,
No.~RS-2024-00354342
and
No.~RS-2025-02219521,
Radiation Science Research Institute,
Foreign Large-Size Research Facility Application Supporting project,
the Global Science Experimental Data Hub Center, the Korea Institute of Science and
Technology Information (K25L2M2C3 ) and KREONET/GLORIAD;
the Polish Ministry of Science and Higher Education and 
the National Science Center;
the Ministry of Science and Higher Education of the Russian Federation
and the HSE University Basic Research Program, Moscow; % from 15.04.2021
University of Tabuk research grants
S-1440-0321, S-0256-1438, and S-0280-1439 (Saudi Arabia);
the Slovenian Research Agency Grant Nos. J1-50010 and P1-0135;
Ikerbasque, Basque Foundation for Science, and the State Agency for Research
of the Spanish Ministry of Science and Innovation through Grant No. PID2022-136510NB-C33 (Spain);
the Swiss National Science Foundation; 
the Ministry of Education and the National Science and Technology Council of Taiwan;
and the United States Department of Energy and the National Science Foundation.
These acknowledgements are not to be interpreted as an endorsement of any
statement made by any of our institutes, funding agencies, governments, or
their representatives.
We thank the KEKB group for the excellent operation of the
accelerator; the KEK cryogenics group for the efficient
operation of the solenoid; and the KEK computer group and the Pacific Northwest National
Laboratory (PNNL) Environmental Molecular Sciences Laboratory (EMSL)
computing group for strong computing support; and the National
Institute of Informatics, and Science Information NETwork 6 (SINET6) for
valuable network support.

\bibliography{main}

\clearpage
\onecolumngrid
\appendix
\counterwithin{figure}{section}

\section*{END MATTER}
\renewcommand{\thefigure}{\arabic{figure}}
\setcounter{figure}{2} 
The decay width for the $\Bp\to\Kp\alp$ production process is given by Ref. \cite{Izaguirre_2017} as $\Gamma(\Bp\to\Kp\alp) = m^3_B|g_{abs}|^2(1 - m^2_K/m^2_B)^2 f^2_0(m_a^2)\lambda^{1/2}_{Ka}/64\pi$, with $f_0(m_a^2) = 0.330/(1 - m_a^2/37.46)$ and $\lambda_{Ka} = (1 - (m_a + m_K)^2/m_B^2)(1 - (m_a - m_K)^2/m_B^2)$, where $f_0(q)$ is a form factor from the hadronic matrix element \cite{PhysRevD.71.014015} and $\lambda$ is the Källen function \cite{Kallen:1964lxa}. 

The ALP coupling to the $b$ and $s$ quarks $g_{abs}$ can be written in terms of the ALP-$W$ coupling $g_{aW}$ as $g_{abs} = -3\sqrt{2}G_Fm_W^2g_{aW}\sum_{q = c,t}V_{q b}V^*_{q s}f(m_q^2/m_W^2)/16\pi^2$, where $f(x) = x[1 + x(\log x -1)]/(1 - x^2)$. Here, $G_F$ is the Fermi constant, $V_{q s}$ with $q = c,t$ are CKM matrix elements, and $m_B$, $m_K$, $m_W$, $m_\alp$, $m_q$ are the masses of the $\Bp$ and $\Kp$ mesons, $W$ boson, ALP, and $c$ or $t$ quark, respectively. Analogous expressions apply for $\Bp\to\pip\alp$. 

\begin{figure*}[htb]
\includegraphics[width=\textwidth]{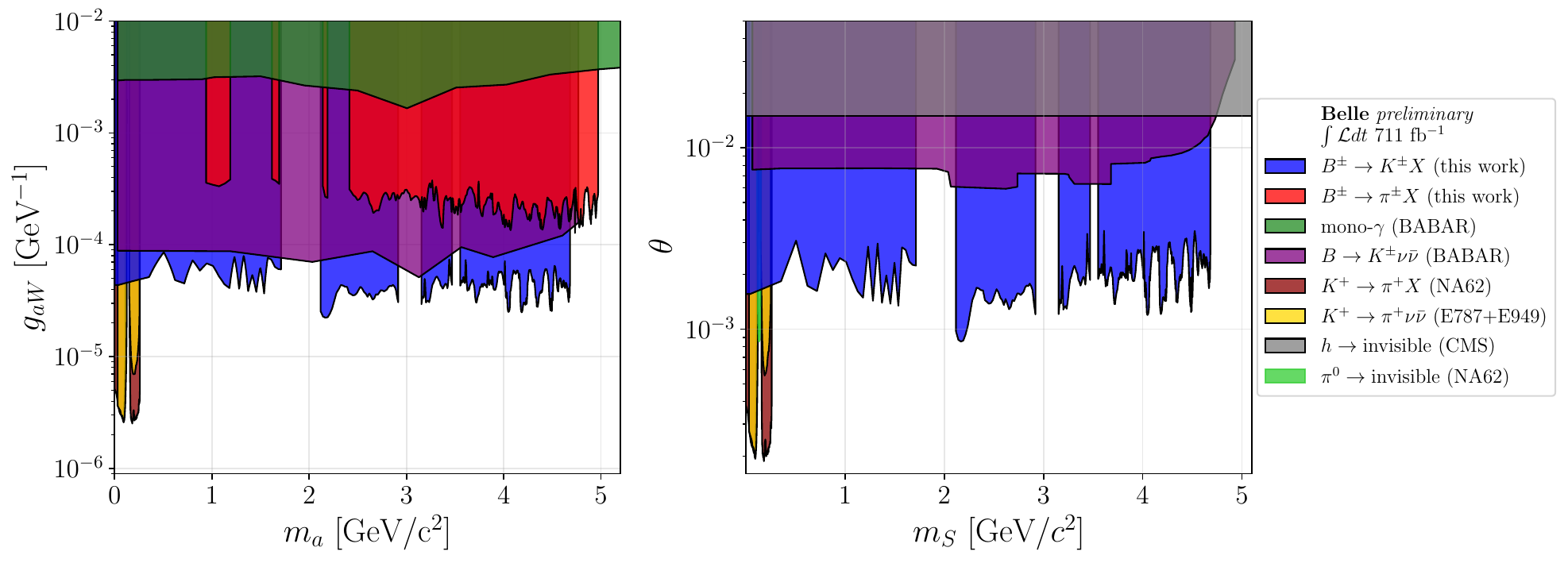}
\caption{\label{fig:fig3}Excluded regions for $m_{\Xinv}$ up to 5.1 \gevcc at 90\% CL shown for: (a) the coupling $g_{aW}$ in the invisible ALP scenario from Ref. \cite{Izaguirre_2017}, and (b) the mixing angle $\theta$ for the dark scalar in the missing energy scenario from Ref. \cite{PhysRevD.101.095006}. Exclusions from BABAR \cite{Lees_2013,BabarMono}, NA62 \cite{NA62:2020xlg,NA62:2020pwi}, E949 \cite{BNL-E949:2009dza}, and CMS \cite{CMS:2018uag} are overlaid for comparison.}
\end{figure*}

\begin{figure*}[htb]
    \centering
    \begin{minipage}[b]{0.6\textwidth}
        \centering
        \includegraphics[width=\textwidth]{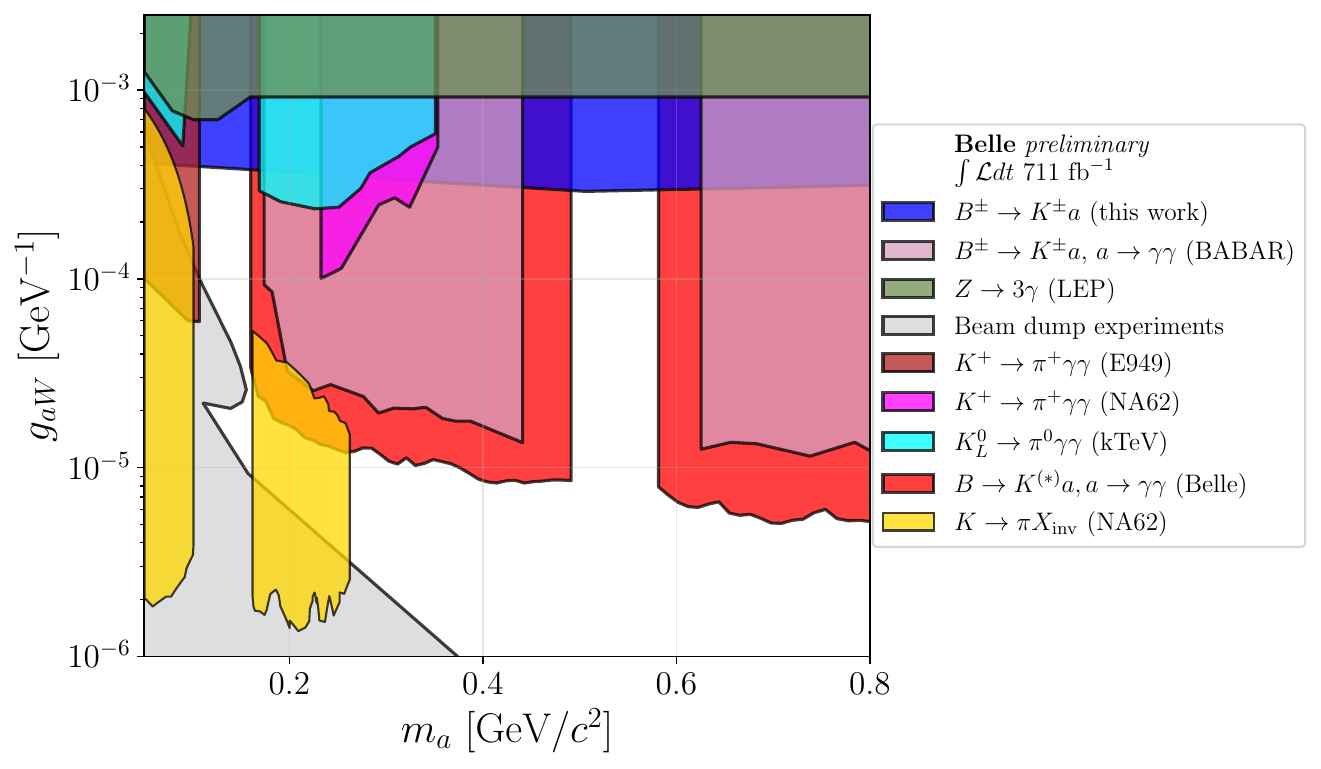}
    \end{minipage}
    \caption{\label{fig:g_aW_Visible} Limits on the coupling strength $g_{aW}$ assuming the ALP decays via $\alp\to\gamma\gamma$, including the constraints from other experiments presented in Ref. \cite{Izaguirre_2017}, with the addition of more recent constraints that appear in Ref. \cite{Sungjin25}.}
\end{figure*}

\Cref{fig:fig3}a shows limits on $g_{aW}$ for invisible ALPs. \Cref{fig:g_aW_Visible} presents the corresponding 90\% CL upper limits assuming $\alp\to\gamma\gamma$. In this case, the decay width and therefore lifetime are related to the coupling via $\Gamma_a = g_{aW}^2m_a^3\sin^4\theta_W/64\pi$, where $\theta_W$ is the Weinberg angle.

\begin{figure*}[htb]
    \centering
    \begin{minipage}[b]{\textwidth}
        \centering
        \includegraphics[width=\textwidth]{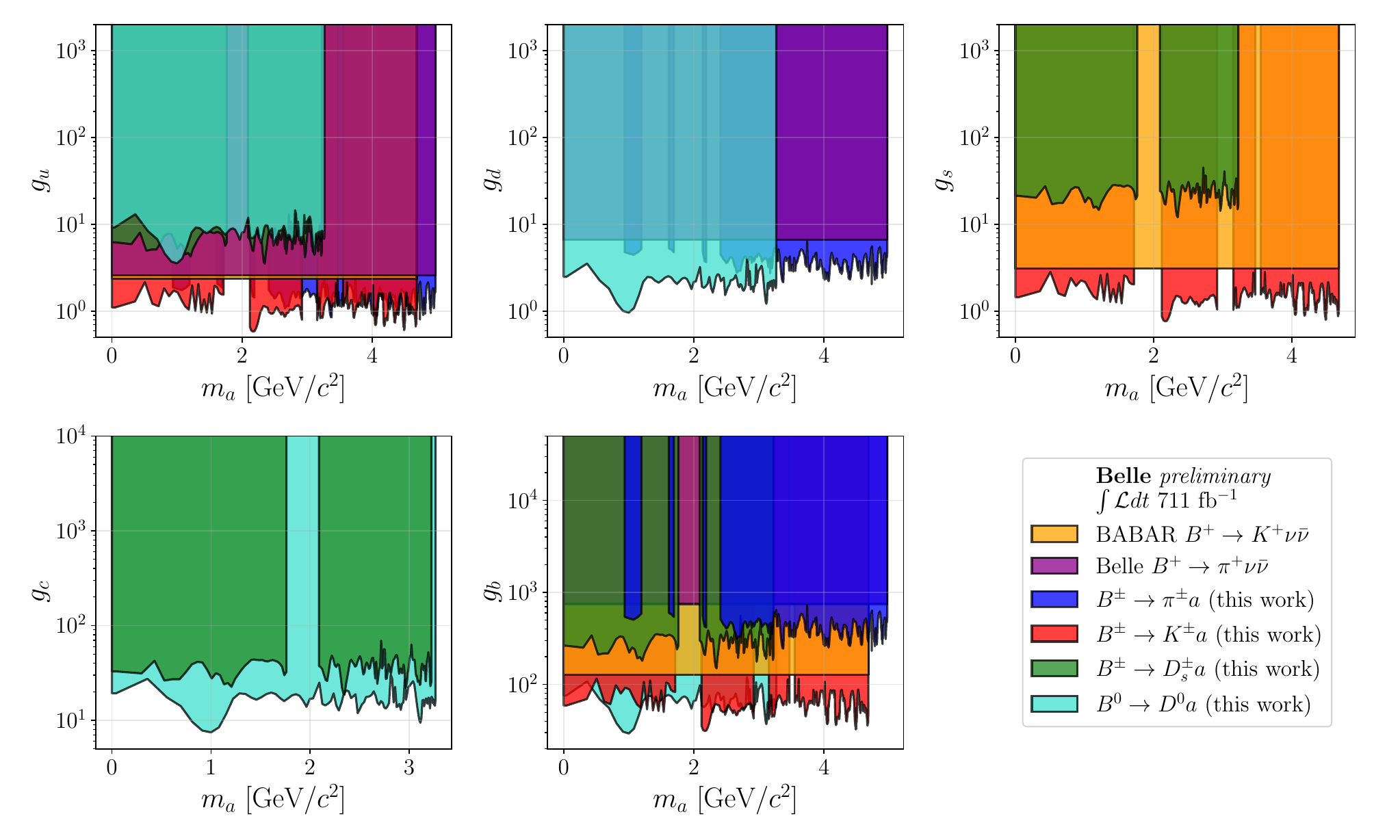}
    \end{minipage}
    \caption{\label{fig:ALPq_couplings} 90\% confidence level exclusion regions on the ALP-quark couplings described in Ref. \cite{zhang2023belle}.}
\end{figure*}

Following Ref. \cite{zhang2023belle}, the exclusion limits can also be mapped to the ALP-quark couplings $g_q$ using the $\mathcal{B}(B\to h\alp)g_q^{-2}$ values given in that reference to translate the branching fraction limits in \cref{fig:BF_SensAll} to constraints on $g_q$. The resulting exclusion regions, shown in \cref{fig:ALPq_couplings}, illustrate the complementary sensitivity of different $B$ decay channels.

Alternatively, one can consider a process that produces a dark scalar $S$ in $B$ meson decays. In Ref. \cite{PhysRevD.101.095006}, the decay width for such a process is given by $\Gamma(\Bp\to\Kp\S) = \sqrt{2}G_F|C_{bs}|^2(m_b + m_s)^2(m^2_B - m^2_K)^2[(m^2_B - m^2_K - m^2_S)^2 - 4m^2_Km^2_s]^{1/2}/[64\pi m^3_B(m_b - m_s)^2]$, where $f_0(q^2)$ is the scalar hadronic form factor from Ref. \cite{PhysRevD.93.025026} and $C_{bs}$ is a Wilson coefficient that can be written as $C_{bs} = 3\sqrt{2}G_Fm_t^2V_{tb}V^*_{tq}\sin\theta/16\pi^2 + \mathcal{O}(m_S^2/m^2_W)$, where $\theta$ is the dark scalar-Higgs mixing angle, and $m_t$ and $m_S$ are the masses of the top quark and dark scalar. 

Whether a dark scalar decays visibly or invisibly depends on both the mixing angle $\theta$ and Yukawa coupling $y_\chi$ to dark fermions $\chi$. The invisible decay rate is written in Ref. \cite{PhysRevD.101.095006} as $\Gamma(\S\to\chi\chi) = y_\chi^2m_S\cos^2\theta(1 - 4m_\chi^2/m_S^2)^{3/2}/8\pi$. Assuming the coupling $y_\chi$ is sufficiently large so that $\mathcal{B}(S\to\chi\chi) \approx 1$, we can map the branching fraction limits in \cref{fig:BF_SensAll} to the mixing angle $\theta$. These results are shown in \cref{fig:fig3}b.

\begin{figure*}[htb]
    \centering
    \begin{minipage}[b]{0.45\textwidth}
        \centering
        \includegraphics[width=\textwidth]{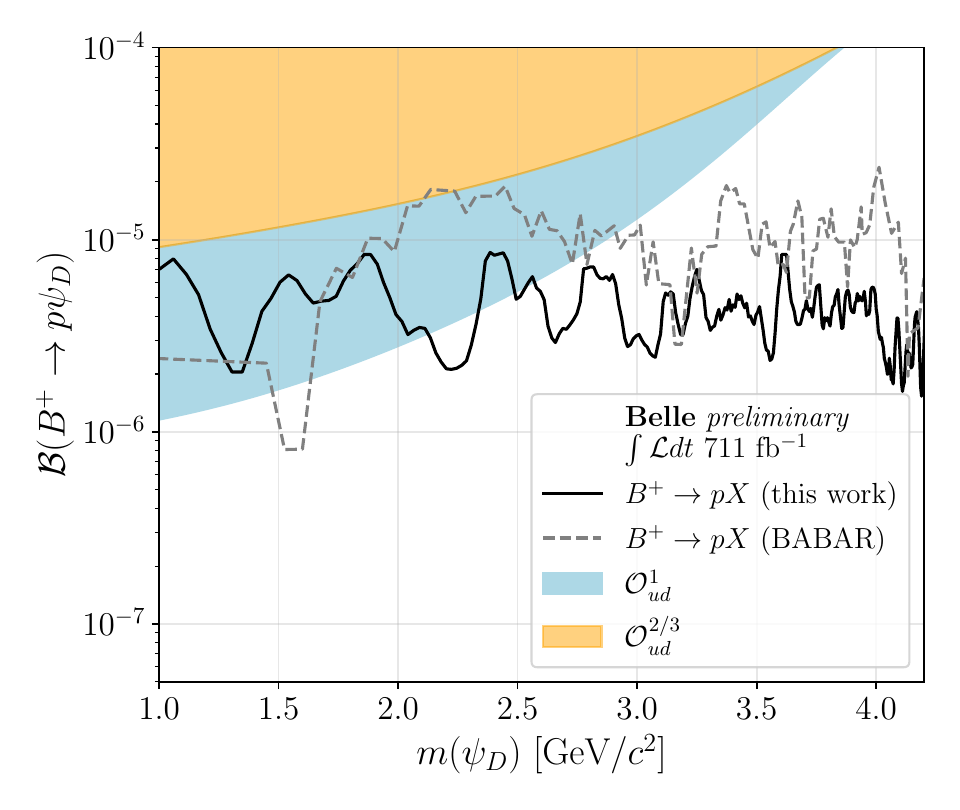}
    \end{minipage}
    \begin{minipage}[b]{0.45\textwidth}
        \centering
        \includegraphics[width=\textwidth]{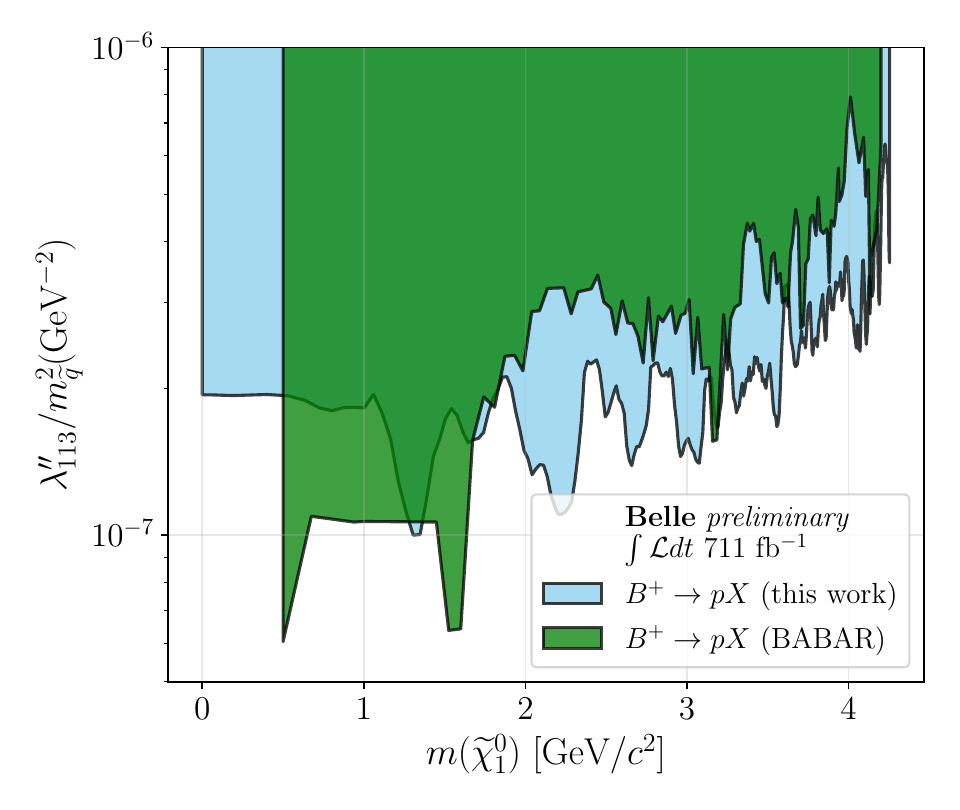}
    \end{minipage}    
    \caption{\label{fig:Bmeso+RPV}Left: Theory expectation for the $B$-mesogenesis effective operators with 90\% CL upper limits on the branching fraction $\Bp\to\pp\psiD$ derived from the Belle dataset. Right: Excluded parameter space by BABAR and Belle (this work) for the RPV coupling $\lambda^{\prime\prime}_{113}$ as a function of neutralino mass.}
\end{figure*}

$B$-mesogenesis refers to a baryogenesis mechanism via decays of $B$ mesons to dark baryons in the early universe \cite{Elor_2019}. These decays take the form $B\to\mathcal{B}_{ij}\mathcal{M}\psiD$, where $\mathcal{B}$ is a baryon containing quarks $i=u,c$ and $j=d,s$, $\mathcal{M}$ is a meson, and \psiD is the dark baryon. The interactions are described by effective operators $\mathcal{O}_{ij}$ for each possible $\bbar\to ij\psiD$ vertex. As this Letter probes $B$ decays with a proton in the final state, it can only constrain the $\mathcal{O}_{ud}$ operator. For this mechanism to produce the observed baryon asymmetry of the universe, the branching fraction of $\Bp\to\mathcal{B}_{ij}\mathcal{M}\psiD$ must be greater than $10^{-6}$ \cite{Alonso_lvarez_2021}. The relation between the decay rates to $\Bp\to\mathcal{B}_{ij}\mathcal{M}\psiD$ and the two-body decay $\Bp\to\mathcal{B}_{ij}\psiD$ is also described in Ref. \cite{Alonso_lvarez_2021}. The limit of $10^{-6}$ on the multi-body branching fraction is mapped onto the two-body branching fraction, shown in coloured regions in \cref{fig:Bmeso+RPV}. The branching fraction limit on $\Bp\to\pp\X$ in \cref{fig:BF_SensAll} rules out $B$-mesogenesis for dark baryon masses above 2.5 \gevcc, and between 2.15 and 2.3 \gevcc.

The R-parity violating (RPV) supersymmetry process $\Bp\to\pp\Tilde{\chi}_0$, where $\Tilde{\chi}_0$ is the lightest neutralino, can also be constrained by the limits on $\Bp\to\pp\X$. The neutralino production process depends on the RPV coupling $\lambda^{\prime\prime}_{113}$ divided by a squared squark mass $m^2_{\Tilde{q}}$. The results for $\Bp\to\pp\X$ in \cref{fig:BF_SensAll} are mapped onto this coupling as a function of neutralino mass using results from \cite{Dib_2023}. We exclude lower values of the coupling than the constraints from BABAR between 0--0.5 $\gevcc$, and for most masses above 2 $\gevcc$.

\clearpage
\appendix
\counterwithin{figure}{section}

\section*{SUPPLEMENTAL MATERIAL}
%\section{Axion-like particle couplings}
\renewcommand{\thefigure}{\arabic{figure}}
\setcounter{figure}{6}  % Reset figure numbering for the supplement
%\section{Analysis  method and validation}
The mass point with the most significant excess occurred in the search for $\Bp\to\pip\Xinv$ at $m_{\Xinv} = 3.28 \gevcc$. The best fit result at this scan point is shown in \cref{fig:fitResult}. The full $p_h$ distributions are shown in \cref{fig:p_distributions2}, with simulated events overlaid. The simulated events are reweighted to match the Belle dataset's luminosity, as well as the differences in reconstruction efficiency between real and simulated events.
The analysis reconstruction efficiency and BDT responses were validated on a sideband samples containing the control modes $\Bp\to\jpsi\Kp$ and $\Bz\to\Dzb\piz$ prior to unblinding the signal region data. The outputs of the background suppression BDTs are shown for both control modes in \cref{fig:BDT_responses}.

\begin{figure}[hbt]
    \centering
    \begin{minipage}[b]{0.75\textwidth}
    \includegraphics[width=\textwidth]{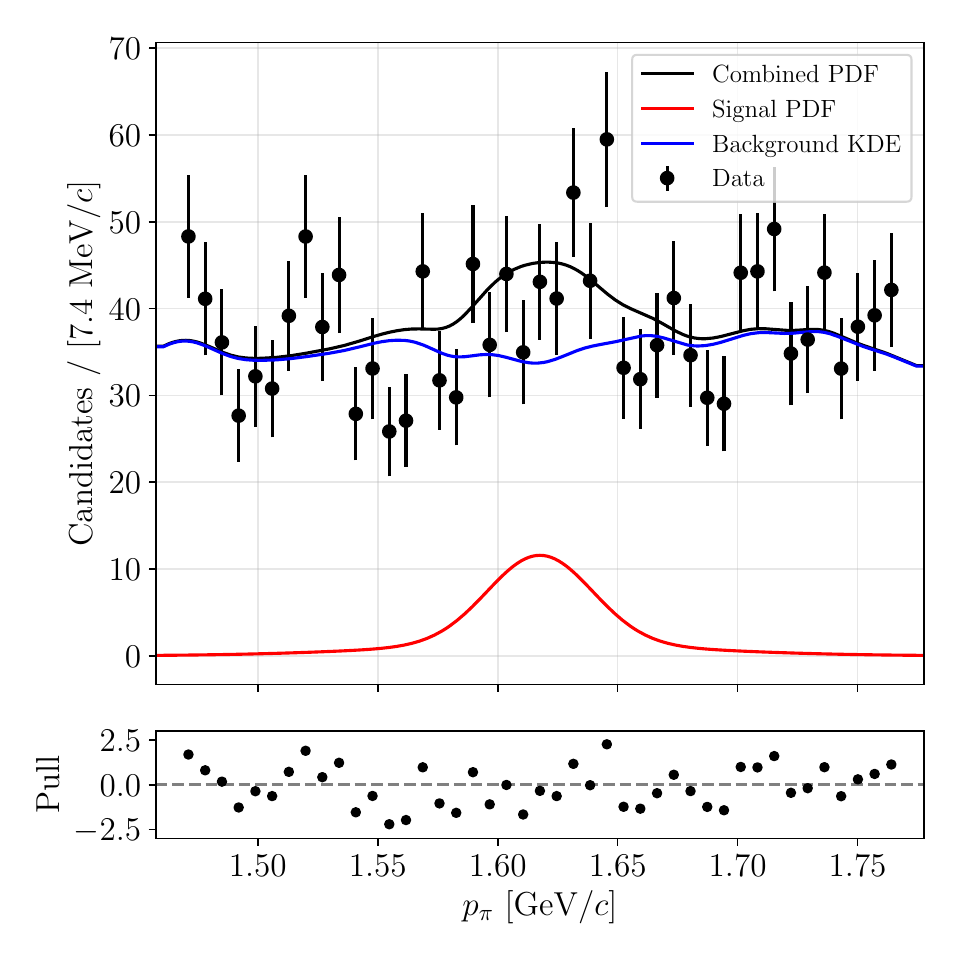}
    \end{minipage}
    \caption{\label{fig:fitResult} Signal extraction fit with the highest local significance, corresponding to $m_{\Xinv} = 3.28 \gevcc$ for the channel $\Bp\to\pip\Xinv$.}
\end{figure}

\begin{figure*}[htb]
    \centering
    \begin{minipage}[b]{0.4\textwidth}
        \centering
        \includegraphics[width=\textwidth]{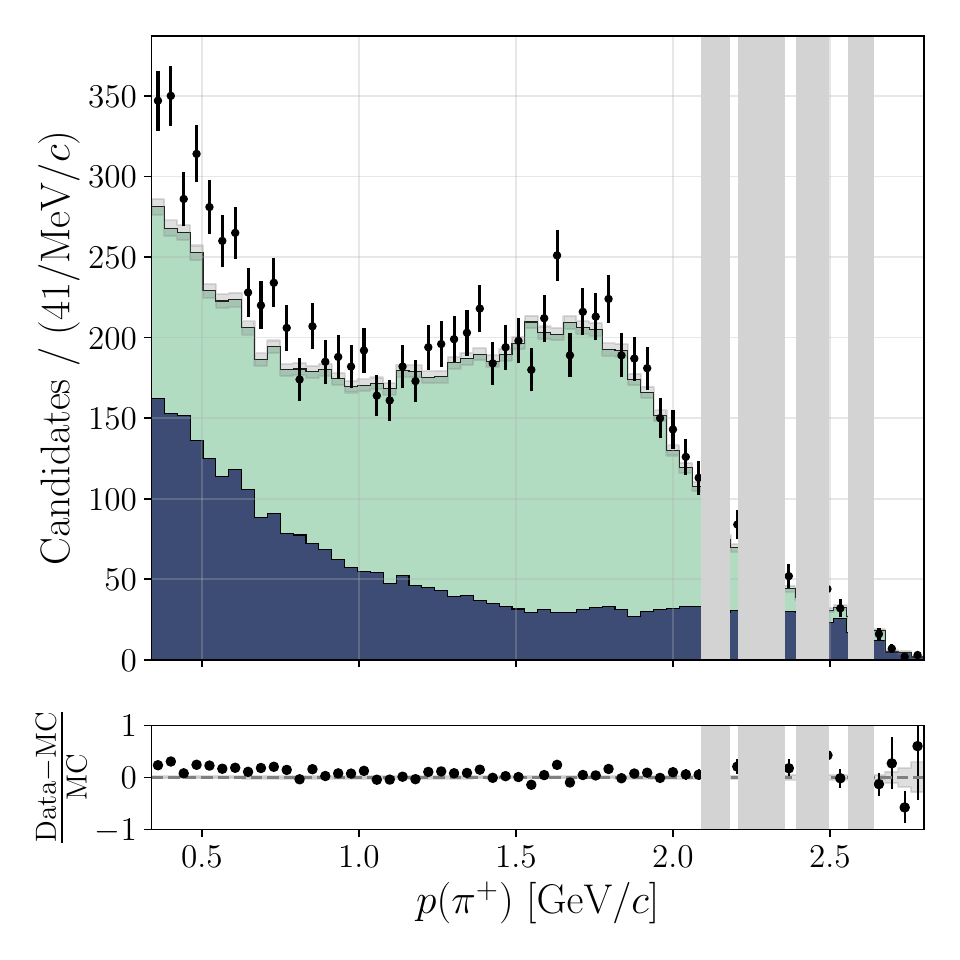}
    \end{minipage}
    \begin{minipage}[b]{0.4\textwidth}
        \centering
        \includegraphics[width=\textwidth]{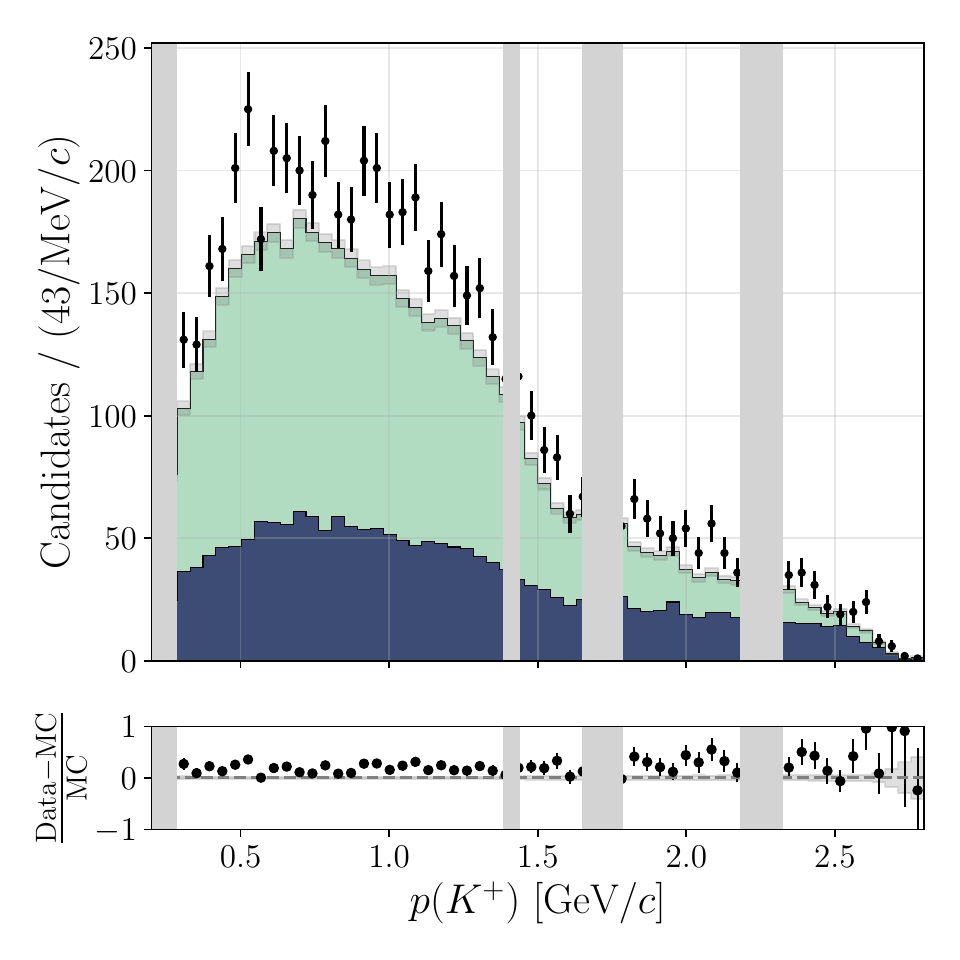}
    \end{minipage}  
    \begin{minipage}[b]{0.4\textwidth}
        \centering
        \includegraphics[width=\textwidth]{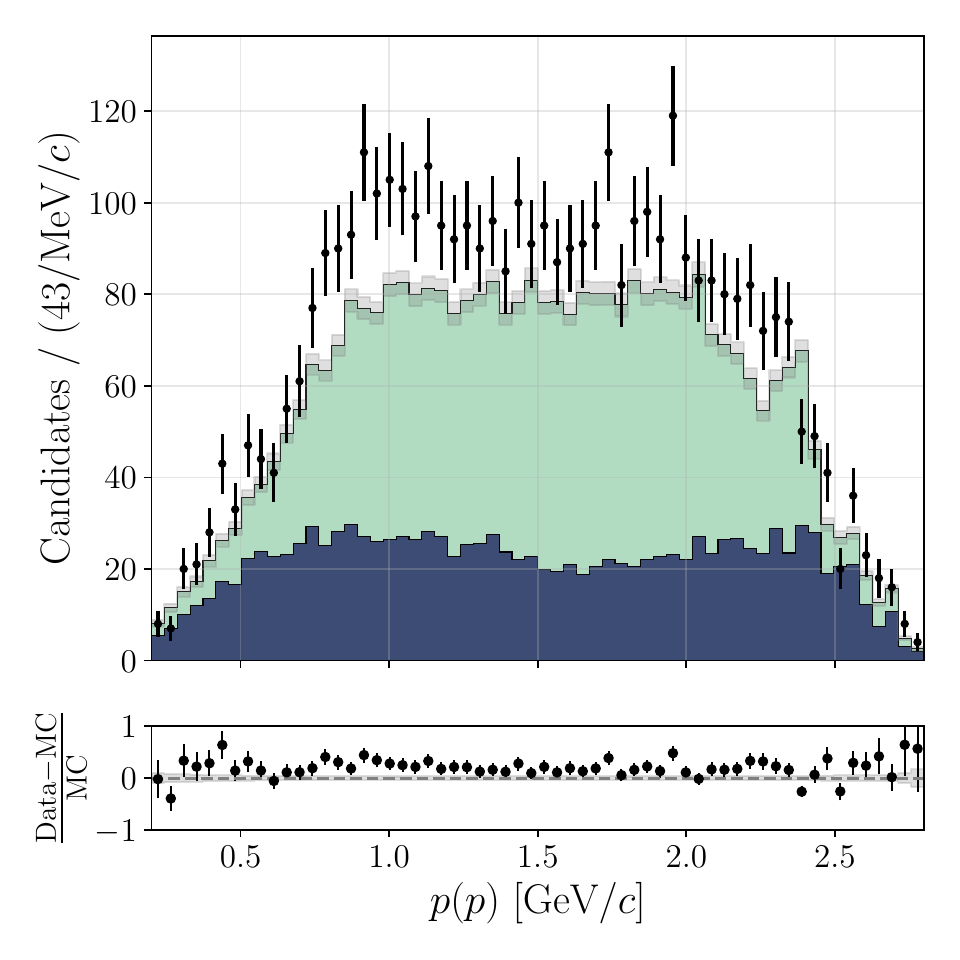}
    \end{minipage}
    \begin{minipage}[b]{0.4\textwidth}
        \centering
        \includegraphics[width=\textwidth]{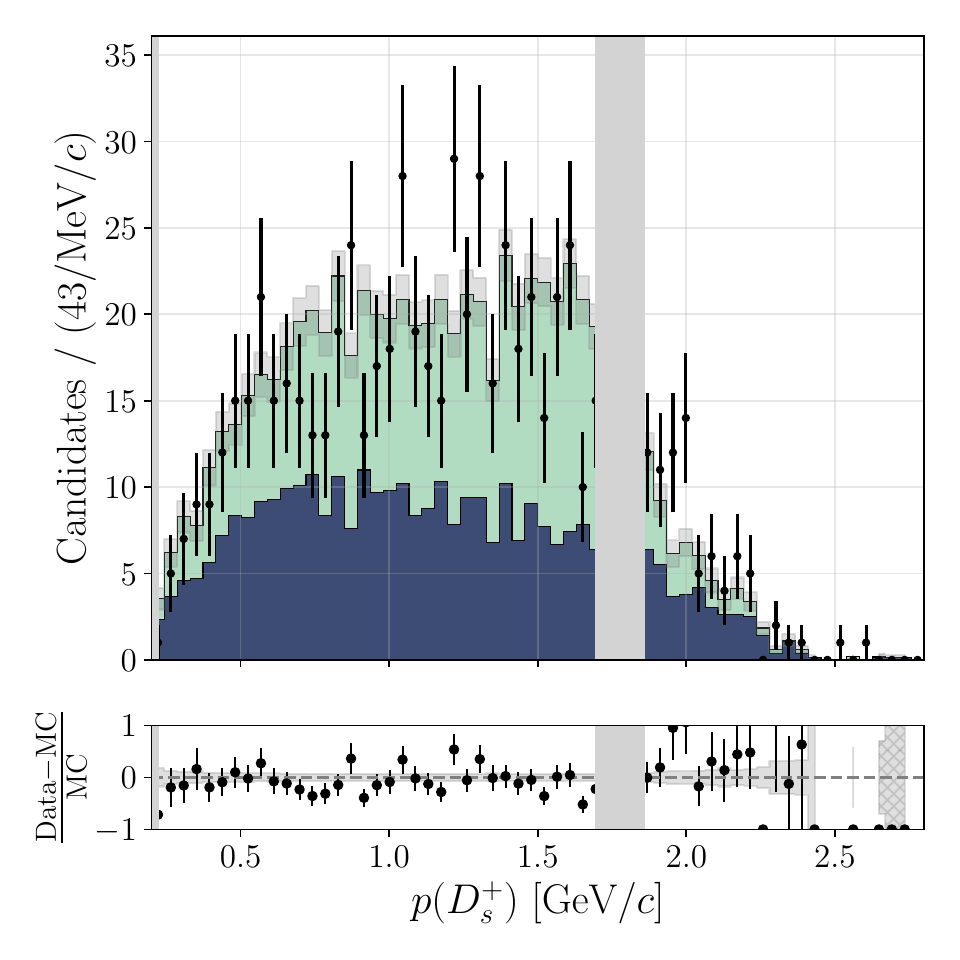}
    \end{minipage}  
    \begin{minipage}[b]{0.4\textwidth}
        \centering
        \includegraphics[width=\textwidth]{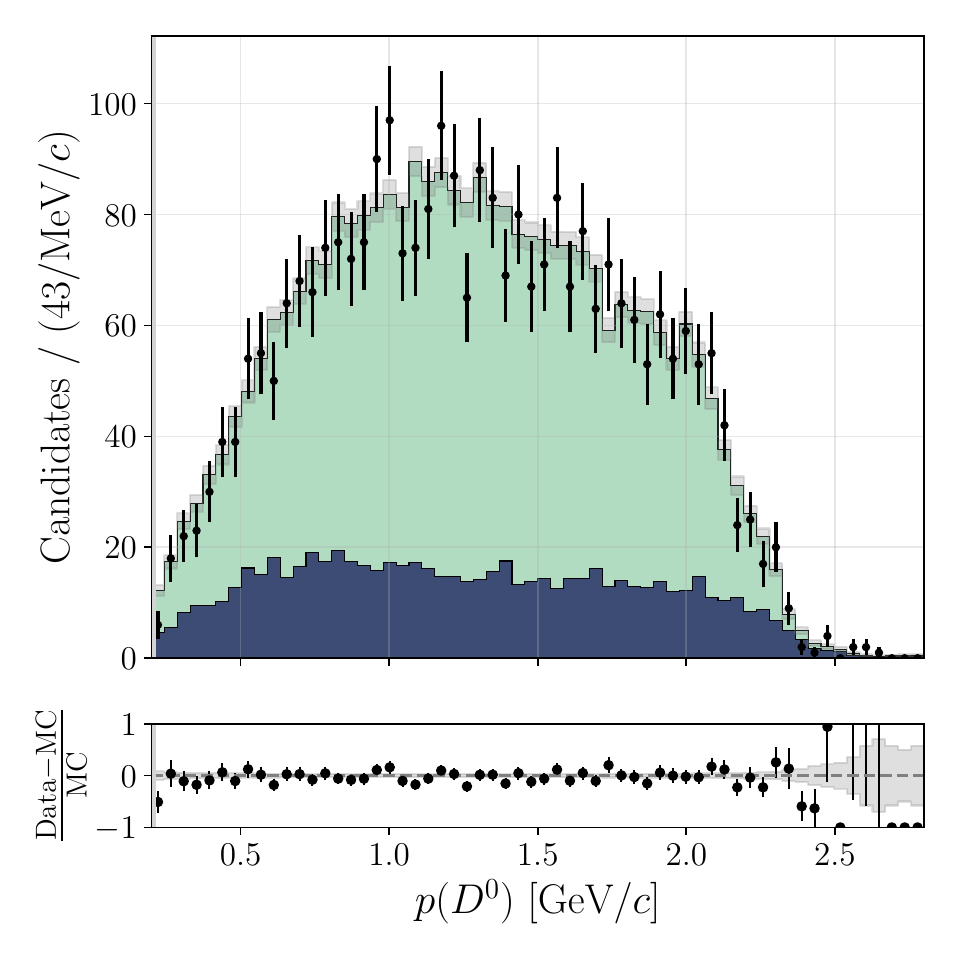}
    \end{minipage}  
    \begin{minipage}[t]{0.4\textwidth}
        \centering
        \vspace{-5cm}
        \includegraphics[width=0.6\textwidth]{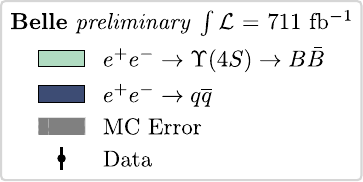}
    \end{minipage}  
    \caption{\label{fig:p_distributions2} Full $p_h$ distributions for the five search channels. Grey bands indicate regions of phase space vetoed in the signal extraction. The overlaid simulated events are reweighted to match the Belle dataset integrated luminosity, and to account for the discrepancy in reconstruction efficiency arising from the FEI.}
\end{figure*}

\begin{figure}[htb]
    \centering
    \begin{minipage}[b]{0.45\textwidth}
        \centering
        \includegraphics[width=\textwidth]{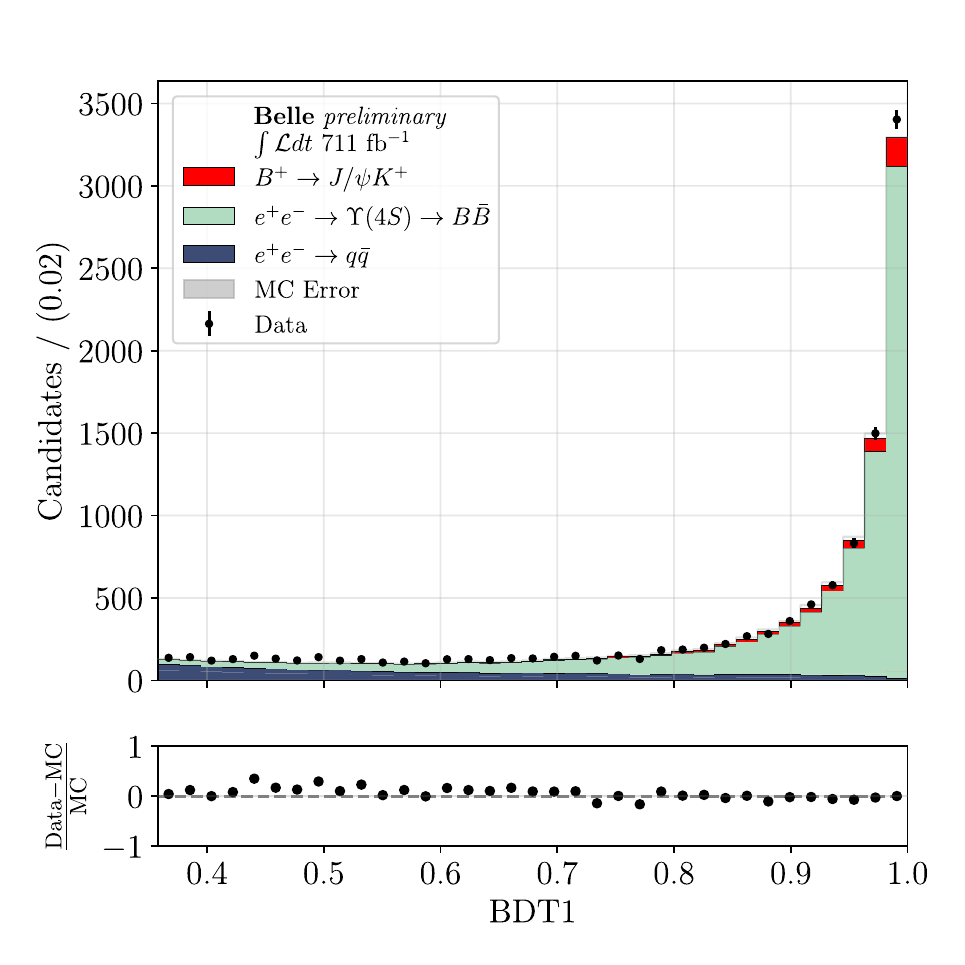}
    \end{minipage}
    \begin{minipage}[b]{0.45\textwidth}
        \centering
        \includegraphics[width=\textwidth]{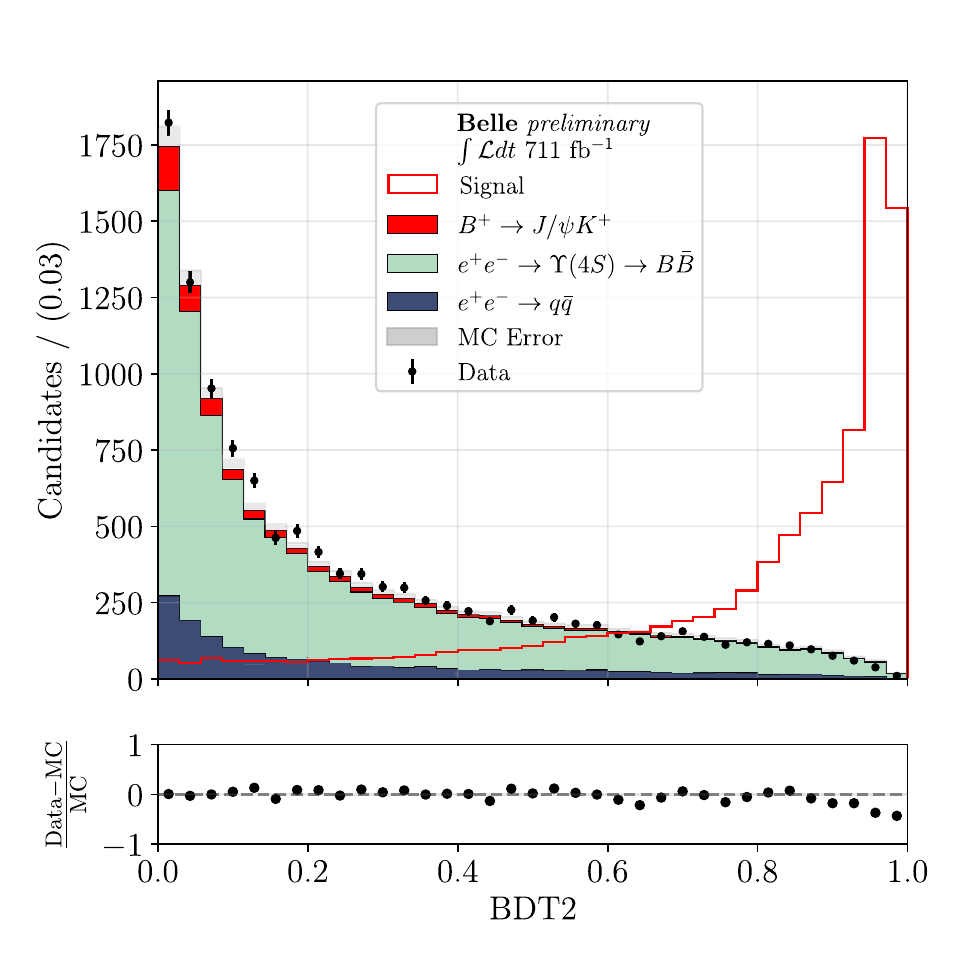}
    \end{minipage}  \\
    \begin{minipage}[b]{0.45\textwidth}
        \centering
        \includegraphics[width=\textwidth]{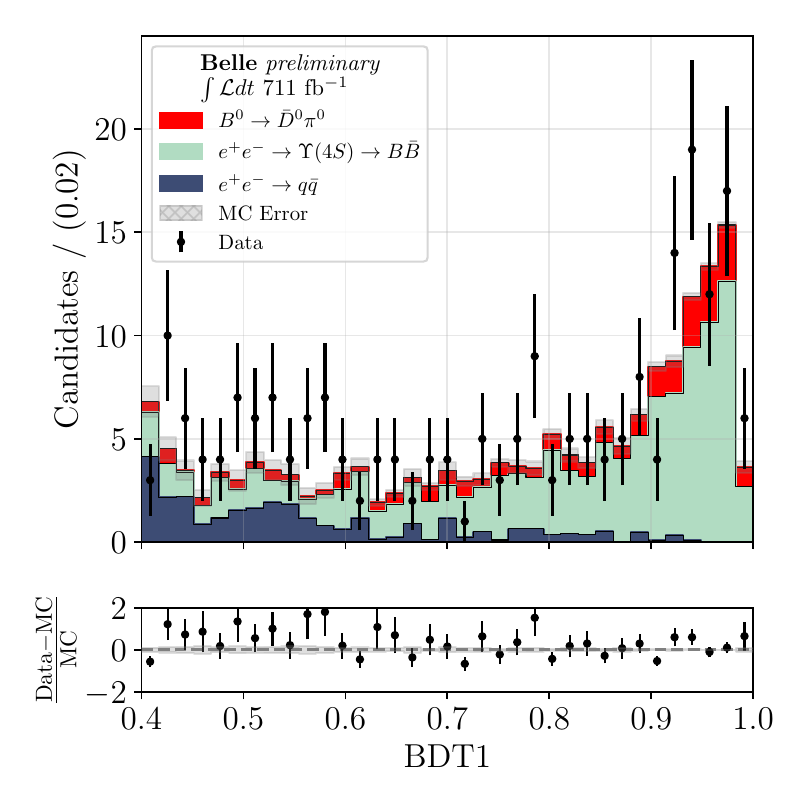}
    \end{minipage}
    \begin{minipage}[b]{0.45\textwidth}
        \centering
        \includegraphics[width=\textwidth]{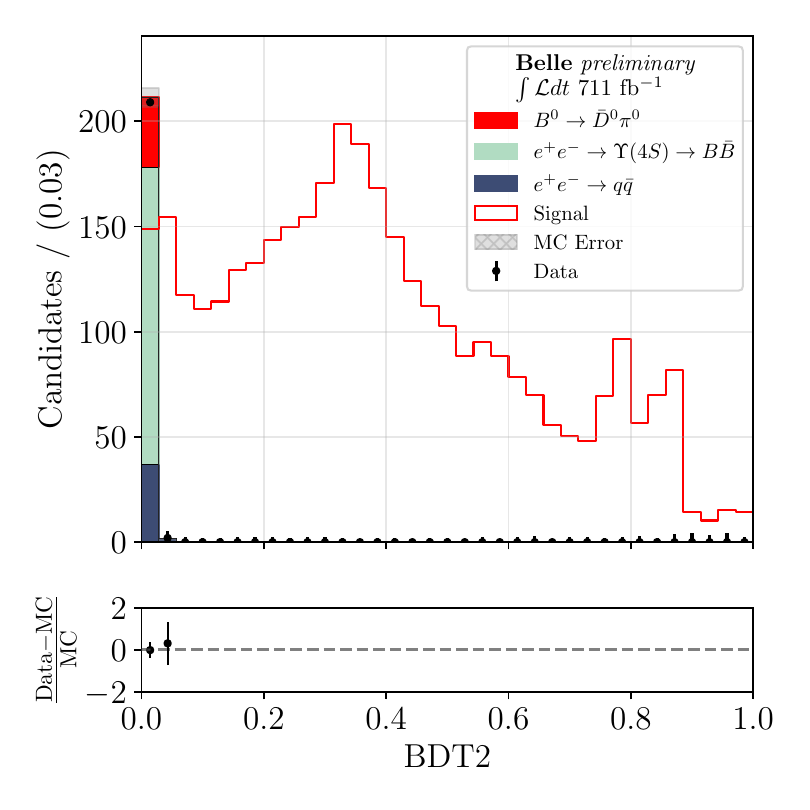}
    \end{minipage}  
    \caption{\label{fig:BDT_responses} BDT response comparison between simulated and real data in the two-track and $\eecl$ sidebands, which contain the $\Bp\to\jpsi\Kp$ (top row) and $\Bz\to\Dzb\piz$ (bottom row) control modes respectively, used to validate the performance of the BDTs.}
\end{figure}

\end{document}